\documentstyle[12pt]{article}

\def\tend{\mathop{\to}}

\def\lim{\mathop{\rm {lim}}}
\def\be{\begin{equation}}
\def\ee{\end{equation}}
\begin{document}
\large
\rm

\vskip 2cm

\begin{center}
\bigskip
{\bf Ultraviolet divergences, renormalization and nonlocality
of interactions in quantum field theory.}\\
\bigskip
{\rm
Renat Kh.Gainutdinov}\\\bigskip
\it
Department of Physics\\
Kazan State University,\\
18 Kremlevskaya St, Kazan 420008,\\
Russia\\
E-mail: Renat.Gainutdinov@ksu.ru
\end{center}
\section*{\bf Abstract}
\vskip 1.5em

\parindent=2.5em

{\rm\par

We discuss the dynamical situation which arises in a local quantum
field theory after renormalization. By using the example
of the three-dimensional theory of a neutral scalar field
interacting through the quartic coupling, we show that after 
renormalization the dynamics of a theory is governed by a
generalized dynamical equation with a nonlocal interaction operator. 
It is shown that the generalized dynamical equation allows one
to formulate this theory in an ultraviolet-finite way.

PACS numbers: 03.70.+k, 11.10.Gh., 11.10.Kk

\newpage
\parindent=2.5em

The problem of the ultraviolet (UV) divergences is
one of the most important problems of quantum field theory (QFT).
The renormalized program has not resolved this problem as a whole.
For example, regularization of the the scattering matrix gives rise
to the situation in which divergent terms automatically appear in the
Schr{\"o}dinger and Tomonago-Schwinger equations. For this reason
these equations are only of formal importance to QFT, and the 
question of what kind of equation governs the dynamics of a theory
after renormalization remains unanswered.
Since locality has been argued to be the main cause of
infinities in QFT, it seems natural to resolve this problem by introducing
a nonlocal form factor into the interaction Hamiltonian or
Lagrangian of a quantum field theory (see, for example, Ref.[1] and references
therein). However, this way of nonlocalization gives rise to several conceptual
problems, since such an introduction of a nonlocal form factor is not
intrinsically consistent. In fact, the Schr{\"o}dinger equation 
is local-in-time, and interaction
Hamiltonians describe an instantaneous interaction. In nonrelativistic
quantum mechanics processes of instantaneous interaction may be nonlocal in
space. But in relativistic quantum theory a local-in-time process must also
be local in space. Thus, for nonlocalization to be
intrinsically consistent, Hamiltonian dynamics need to be extended to describe
the evolution of a quantum system whose dynamics is generated by a
nonlocal-in-time interaction. This problem has been solved in Ref.[2], where
it has been shown that the current concepts of quantum physics allow
such an extension of quantum dynamics. As a consequence of the general
postulates of the canonical and Feynman approach to quantum theory
a generalized dynamical equation has been derived. 
Being equivalent to the Schr{\"o}dinger equation in the case of 
instantaneous interactions, this equation
allows the generalization to the case of nonlocal-in-time interactions.
This has been shown [2] to open new possibilities to resolve the problem 
of the UV divergences in QFT.

Note that the problem of the UV divergences and renormalization is
very significant  even for describing nucleon dynamics at low energies.
In fact, Lagrangians of effective field theories (EFT's) [3] that become
very popular in nuclear physics [4] contain terms leading to
the UV divergences, and  renormalization
is needed. A fundamental difficulty in an EFT description of
nuclear forces is that are nonperturbative, and one has to
renormalize the Schr{\"o}dinger and Lippmann-Schwinger (LS) equations
for describing low-energy nucleon dynamics. However, renormalization
gives rise to singular potentials in the case of which these
equations have no sense. 
At the same time, being a unique consequence of the most general
principles on which quantum theory is founded, 
the generalized dynamical equation must be 
satisfied in all the cases. In Ref.[5] it has been shown 
that the T matrix obtained in Ref.[6], by 
renormalizing a toy model of the NN interaction, does not
satisfy the LS equation but satisfies the generalized dynamical
equation with a nonlocal-in-time interaction operator. This
gives reason to suppose that in any theory with the UV divergences
after regularization and renormalization the dynamics of the
theory is governed by the generalized dynamical equation with
such interaction operators. In the present paper the truth of
the above assumption is proved by using the example of the supernormalizable
theory $\varphi_3^4$. We show that after renormalization the dynamics
of the theory is governed by the generalized dynamical equation with
nonlocal interaction operator, and within a generalized quantum
dynamics(GQD) developed in Ref.[2] this theory can be formulated
in an ultraviolet-finite way.

In the GQD the following assumptions are used as basic
postulates:

(i) The physical state of a system is represented by a vector
(properly by a ray) of a Hilbert space.

(ii) An observable A is represented by a Hermitian hypermaximal operator
$\alpha$. The eigenvalues $a_r$ of $\alpha$ give the possible values of A.
An eigenvector $|\varphi_r^{(s)}>$ corresponding to the eigenvalue
$a_r$ represents a state in which A has the value $a_r$. If the system
is in the state $|\psi>,$ the probability $P_r$ of finding the value
$a_r$ for A, when a measurement is performed, is given by
$$P_{r} = <\psi|P_{V_{r}} |\psi>= \sum_s |<\varphi_r^{(s)}|\psi>|^2, $$
where $P_{V_{r}}$ is the projection operator on the eigenmanifold $V_r$
corresponding to $a_r,$ and the sum $\Sigma_s$ is taken over a complete
orthonormal set ${|\varphi_r^{(s)}>}$ (s=1,2,...) of $V_r.$
The state of the system immediately after the observation is
described by the vector $P_{V_{r}}|\psi>.$

These  assumptions are the main assumptions on which quantum
theory is founded.
In the canonical formalism these postulates
are used together with the assumption that the time
evolution of a state vector is governed by the
Schr{\"o}dinger equation.
In the formalism [2] this assumption is not used.
Instead the assumptions (i) and (ii) are used together
with the following postulate.

(iii) The probability of an event is the absolute square of a
complex number called the probability amplitude.
The joint probability amplitude of a time-ordered
sequence of events is product of the separate probability
amplitudes of each of these events.
The probability amplitude of an event which can happen in several
different ways is a sum of the probability
amplitudes for each of these ways.

The statements of the assumption (iii) express the well-known law
for the quantum-mechanical probabilities. Within the canonical
formalism this law is derived as one of the consequences of the
theory. However, in the Feynman formulation of quantum theory [7] this
law is directly derived starting from the analysis of the
phenomenon of quantum interference, and is used as a basic
postulate of the  theory.

It is also used the assumption that the time evolution of a quantum system
is described by the evolution equation $|\Psi(t)>=U(t,t_0)|\Psi(t_0)>,$
where $U(t,t_0)$ is the unitary evolution operator
\begin{equation}
U^{+}(t,t_0) U(t,t_0) = U(t,t_0) U^{+}(t,t_0) = {\bf 1},
\end{equation}
with the group property
$U(t,t') U(t',t_0) = U(t,t_0), \quad  U(t_0,t_0) ={\bf 1}$.
Here we use the interaction picture.
According to the assumption (iii), the probability amplitude of an
event which can happen in several different ways is a
sum of contributions from each alternative way.
In particular, the amplitude
 $<\psi_2| U(t,t_0)|\psi_1>$ can be represented as a sum
of contributions from all alternative
ways of realization of the corresponding evolution process.
Dividing these alternatives in different classes, we can then analyze
such a probability amplitude in different ways.
For example, subprocesses with definite instants of the
beginning and  end of the interaction in the system
can be considered as such alternatives.
In this way the amplitude
$<\psi_2|U(t,t_0)|\psi_1>$  can be written in the form [2]
\begin{equation}
<\psi_2| U(t,t_0)|\psi_1> = <\psi_2|\psi_1> +
\int_{t_0}^t dt_2 \int_{t_0}^{t_2} dt_1
<\psi_2|\tilde S(t_2,t_1)|\psi_1>,
\end{equation}
where $<\psi_2|\tilde S(t_2,t_1)|\psi_1>$ is the probability
amplitude that if at time $t_1$ the system was in the state
$|\psi_1>,$ then the interaction in the system will begin at time
$t_1$ and will end at  time $t_2,$ and at this time the system
will be in the state $|\psi_2>.$ Note that in general $\tilde
S(t_2,t_1)$  may be only an operator-valued generalized function
of $t_1$ and $t_2$ [2], since only $U(t,t_0)={\bf 1}+
\int^{t}_{t_0} dt_2 \int^{t_2}_{t_0}dt_1\tilde S(t_2,t_1)$  must
be an operator on the Hilbert space. Nevertheless, it is
convenient to call $\tilde S(t_2,t_1)$ an "operator", using this
word in generalized sense. In the case of an isolated system the
operator $\tilde S(t_2,t_1)$ can be represented in the form
$\tilde S(t_2,t_1) = exp(iH_0t_2) \tilde T(t_2-t_1) exp(-iH_0
t_1),$ $H_0$ being the free Hamiltonian [2].

As has been shown in Ref.[2], for the evolution operator $U(t,t_0)$ given
by (2) to be unitary
for any times $t_0$ and $t$,
the operator $\tilde S(t_2,t_1)$ must satisfy the following
equation:
\begin{equation}
(t_2-t_1) \tilde S(t_2,t_1) =
\int^{t_2}_{t_1} dt_4 \int^{t_4}_{t_1}dt_3
(t_4-t_3) \tilde S(t_2,t_4) \tilde S(t_3,t_1).
\end{equation}
This equation allows one to obtain the operators $\tilde S(t_2,t_1)$
for any $t_1$ and $t_2$, if the operators
$\tilde S(t'_2, t'_1)$ corresponding to infinitesimal
duration times $\tau = t'_2 -t'_1$ of interaction are known.
It is natural to assume that most of the
contribution to the evolution operator
in the limit $t_2 \to t_1$ comes from the processes
associated with the fundamental interaction
in the system under study.
Denoting this contribution by
$H_{int}(t_2,t_1)$, we can write
\begin{equation}
\tilde{S}(t_2,t_1) \tend\limits_{t_2\rightarrow t_1}
H_{int}(t_2,t_1) + o(\tau^{\epsilon}),
\end{equation}
where $\tau=t_2-t_1$.
The parameter $\varepsilon$ is determined by demanding that
$H_{int}(t_2,t_1)$ must be so close to the solution of Eq.(3) in
the limit $t_2\tend t_1$ that this equation has a unique solution
having the behavior (4) near the point $t_2=t_1$.Thus this
operator must satisfy the condition
\begin{equation}
(t_2-t_1) H_{int}(t_2,t_1)\tend\limits_{t_2 \tend t_1}
\int^{t_2}_{t_1} dt_4 \int^{t_4}_{t_1} dt_3
(t_4-t_3) H_{int}(t_2,t_4) H_{int}(t_3,t_1)+ o(\tau^{\epsilon+1}).
\end{equation}
Note that the value of the parameter $\epsilon$
depends on the form of the operator $ H_{int}(t_2,t_1).$
Since $\tilde S(t_2,t_1)$ and $H_{int}(t_2,t_1)$ are only
operator-valued distributions, the mathematical meaning of the
conditions (4) and (5) needs to be clarified. We will assume that
the condition (5) means that $<\Psi_2|\int^{t}_{t_0} dt_2
\int^{t_2}_{t_0}dt_1\tilde S(t_2,t_1) |\Psi_1>\tend\limits_{t\tend
t_0}<\Psi_2|\int^{t}_{t_0} dt_2 \int^{t_2}_{t_0}dt_1
H_{int}(t_2,t_1)|\Psi_1>+o(\tau^{\epsilon+2}),$ for any vectors
$|\Psi_1>$ and $|\Psi_2>$ of the Hilbert space. The condition (6)
has to be  considered in the same sense.
 
Within the GQD the operator $H_{int}(t_2,t_1)$ plays the role 
which the interaction Hamiltonian plays in the ordinary formulation
of quantum theory: It generates the dynamics of a system. Being a
generalization of the interaction Hamiltonian, this
operator is called the generalized interaction operator.
If  $H_{int}(t_2,t_1)$ is specified, Eq.(3) allows one to find the
operator $\tilde S(t_2,t_1).$
Formula (2) can then be used to construct the evolution operator
$U(t,t_0)$ and accordingly the state vector
$|\psi(t)> = |\psi(t_0)> +  \int_{t_0}^t dt_2
\int_{t_0}^{t_2} dt_1 \tilde S(t_2,t_1) |\psi(t_0)> $
at any time $t.$ Thus Eq.(3) can be regarded as an equation of motion for
states of a quantum system. 
It should be noted that Eq.(3) is written only in terms of the
operators $\tilde S(t_2,t_1),$ and does not contain operators
describing the interaction in a quantum system. It is a relation
for $\tilde S(t_2,t_1)$ which are the contributions to the evolution
operator from the processes with defined instants of the
beginning and end of the interaction in the system. This relation
is a unique consequence of the unitarity condition (1) and the
representation (2) expressing the Feynman superposition principle
(the assumption (iii)). For this reason the relation (3) must be
satisfied in all the cases. A remarkable feature of this fundamental
relation is that it works as a recurrence relation, and to construct
the evolution operator it is enough to now the contributions to this
operator from the processes with infinitesimal duration times of
interaction, i.e. from the processes of a fundamental interaction
in the system. This makes it possible to use the fundamental
relation (3) as a dynamical equation. Its form does not depend on
the specific features of the interaction ( the Schr{\"o}dinger
equation, for example, contains the interaction Hamiltonian).
Since Eq.(3) must be satisfied in all the cases, it can be considered
as the most general dynamical equation consistent with the current concepts
of quantum theory. As has been shown in Ref.[2], the dynamics
governed by Eq.(3)  is equivalent to the Hamiltonian dynamics in the 
case where the generalized interaction operator is of the form
\begin{equation}
 H_{int}(t_2,t_1) = - 2i \delta(t_2-t_1)
 H_{I}(t_1) ,
\end{equation}
$H_{I}(t_1)$ being the interaction Hamiltonian in the interaction
picture. In this case the evolution operator given by (2) satisfies the
Schr{\"o}dinger equation. 
The delta function $\delta(\tau)$ in (6) emphasizes the fact
that in this case the fundamental interaction is instantaneous. At
the same time, Eq.(3) permits the generalization to the case where
the interaction generating the dynamics of a system is nonlocal in
time [2]. In Ref.[8] this point was demonstrated on exactly solvable models.
It should be noted that form of the generalized interaction operator
cannot be arbitrary, since it must satisfy the condition (5).
As has been shown [2,8], there is one-to-one correspondence between
nonlocality of interaction and the UV behavior of the matrix elements
of the evolution operator as a function of momenta of particles:
The interaction operator can be nonlocal in time only in the case
where this behavior is "bad", i.e. in a local theory it results 
in UV divergences.

The above gives reason to expect that after renormalization
the dynamics of 
a theory does not governed by the Schr{\"o}dinger equation should
mean that interaction in the renormalized theory is nonlocal-in-time.
This has been illustrated [5] on a toy model of the
NN interaction [6]. Let us now consider this
problem by using the example of the three-dimensional theory of a neutral
scalar field interacting through the $\varphi^4$ coupling. The Hamiltonian
of the theory, with a spatial cutoff, is of the form
\begin{equation}
H(g)=H_0+H_I(g),
\end{equation}
where $H_I(g)=\int d^2xg({\bf x}):\varphi^4({\bf x},t=0):$,
$\varphi(x)$ is the field with mass $m$, and $::$ denotes the normal
ordering. As is well known, introducing a spatial cutoff is needed because
of the problems associated with the Haag's theorem. Let $g$ be a
$C_0^\infty(R^2)$ function, $0\leq g\leq 1$.

Since the use of the local interaction Hamiltonian density ${\cal H}_I(x,g)=
g:\varphi^4(x):$ leads to the UV divergences, some regularization
procedure is also needed. Let the regularized interaction Hamiltonian
be of the form
$$H^{(\Lambda)}_I(g)=\int d^2x{\cal H}_I({\bf x},t=0;g,\Lambda)$$
with the interaction Hamiltonian density ${\cal H}_I({\bf x},g,\Lambda)=
g:\varphi_h^4(x):$, where $\varphi_h({\bf x},t)=\int d{\bf x}' h_\Lambda
({\bf x}-{\bf x'})\varphi ({\bf x}',t)$. Here $h_\Lambda({\bf x})$ is a
real function such  that $h_\Lambda({\bf x}-
{\bf x}')\tend\limits_{\Lambda\tend\infty}\delta({\bf x}-
{\bf x}')$, and $<0|\varphi_\Lambda({\bf x},t)\varphi_\Lambda({\bf x}',t)|0>$
is bounded at ${\bf x}=
{\bf x}'$.
 In order that after letting
$\Lambda$ to infinity the theory be finite, we have to complement the
interaction Hamiltonian in the interaction picture
$H^{(\Lambda)}_I(t;g)$ by the renormalization counterterms
$$H^{(\Lambda)}_I(t;g)\tend H^{(r)}_I(t;g,\Lambda)=\int d^2x{\cal H}^{(r)}_I
({\bf x},t;g,\Lambda)$$
with ${\cal H}^{(r)}_I(x;g,\Lambda)={\cal H}^{(\Lambda)}_I(x;g)-
\frac{1}{2}\delta m^2(\Lambda)
:\varphi^2(x):-E(\Lambda)$. The counterterms
$E(\Lambda)$ and $\delta m(\Lambda)$
are the contributions to the ground state energy and to the rest mass of a
single  particle respectively.  Since the operator
$H^{(r)}_I(t;g,\Lambda)$ is self adjoint and bounded from below
on Fock space [9], within perturbation theory one can use
the Dyson expansion  for constructing the evolution operator
$$
U(t,t_0)={\bf 1}+\sum_{n=1}^\infty (-i)^n \int_{t_0}^t dt_1\cdots \int_{t_0}^
{t_{n-1}}dt_n H_I(t_1)\cdots H_I(t_n)=$$
\begin{equation}
={\bf 1}+\sum_{n=1}^\infty \frac{(-i)^n}{n!}\int_{t_0}^tdt_1 \cdots \int_{t_0}^
tdt_n T \left (H_I(t_1)\cdots H_I(t_n)\right ),
\end{equation}
where $T$ denotes the time-ordered product. At the same time we
can rewrite Eq.(8) in the form
\begin{equation}
U(t,t_0)={\bf 1}-\int_{t_0}^t dt_2 \int_{t_0}^
{t_{2}}dt_1 \left (
2i\delta(t_2-t_1)H_I(t_1)+ H_I(t_2)U(t_2,t_1)H_I(t_1)\right ).
\end{equation}
Thus, within Hamiltonian dynamics, the evolution operator can be represented
in the form (2) with the following operator $\tilde S(t',t)$:
$$
\tilde S(t',t)=-2i\delta(t'-t)H_I(t)+
$$
\begin{equation}
+\sum_{n=0}^\infty (-i)^{n+2}
\int_{t}^{t'}dt_1 \cdots \int_{t}^
{t_{n-1}}dt_n H_I(t')H_I(t_1)\cdots H_I(t_n)H_I(t).
\end{equation}
It is not difficult to verify that (10) is the solution of Eq.(3)
with the boundary
condition given by (4) and (6). This means that in the case of the generalized
interaction operators of the form (6), the perturbative
solution of Eq.(3) results in the Dyson expansion that within
Hamiltonian formalism represents the formal solution of the
Schr{\"o}dinger equation.

By using (2) and (10), we can construct the evolution operator
corresponding to the interaction Hamiltonian
$H_I^{(r)}(t)$
$$
U_{r,\Lambda}(t,t_0;g) = {\bf 1} +
\int_{t_0}^t dt_2 \int_{t_0}^{t_2} dt_1
\tilde S_{r,\Lambda}(t_2,t_1;g),
$$
where
\begin{equation}
\tilde S_{r,\Lambda}(t_2,t_1;g)=-2i\delta(t_2-t_1)H_I^{(r)}(t_1)+
\tilde S'_{r,\Lambda}(t_2,t_1;g),
\end{equation}
$$
\tilde S'_{r,\Lambda}(t',t;g)=\sum_{n=0}^\infty (-i)^{n+2}
\int_{t}^{t'}dt_1 \cdots \int_{t}^
{t_{n-1}}dt_n H_I^{(r)}(t';g,\Lambda)\times
$$
\begin{equation}
\times H_I^{(r)}(t_1;g,\Lambda)
\cdots H_I^{(r)}(t_n;g,\Lambda)H_I^{(r)}(t;g,\Lambda).
\end{equation}
Then taking $\Lambda$ to infinity, for the evolution operator and the
S matrix, we get
\begin{equation}
U(t,t_0;g) = \lim\limits_{\Lambda\tend\infty}
U_{r,\Lambda}(t,t_0;g)={\bf 1} +
\int_{t_0}^t dt_2 \int_{t_0}^{t_2} dt_1
\tilde S_{ren}(t_2,t_1;g),
\end{equation}
$$
S(g) = {\bf 1} +
\int_{-\infty}^\infty dt_2 \int_{-\infty}^{t_2} dt_1
\tilde S_{ren}(t_2,t_1;g),
$$
with
\begin{equation}
\tilde S_{ren}(t_2,t_1;g)= \lim\limits_{\Lambda\tend\infty}
\tilde S_{r,\Lambda}(t_2,t_1;g).
\end{equation}
Since the operators $\tilde S_{r,\Lambda}(t_2,t_1;g)$ satisfy Eq.(3) for
any $\Lambda$, the operator $\tilde S_{ren}(t_2,t_1;g)$ should
also satisfy this equation. Thus
$\tilde S_{r,\Lambda}(t_2,t_1;g)$ given by (11) is a solution of the
dynamical equation (3). Let us now find the generalized interaction
operator to which this solution  corresponds. Representing the operator
$\tilde S(t_2,t_1)$ in the form $\tilde S(t_2,t_1) = H_{int}(t_2,t_1)+
\tilde S_1(t_2,t_1),$
dynamical equation (3) can be rewritten as the following equation for
the operator  $\tilde S_{1}(t_2,t_1)$:
$$
(t_2-t_1)\tilde S_1(t_2,t_1) = (t_2-t_1)F(t_2,t_1)+
\int^{t_2}_{t_1} dt_4 \int^{t_4}_{t_1}dt_3
(t_4-t_3)\times
$$
\begin{equation}
\times \left( H_{int}(t_2,t_4) \tilde S_1(t_3,t_1)+
\tilde S_1(t_2,t_4)H_{int}(t_3,t_1)+\tilde S_1(t_2,t_4)
\tilde S_1(t_3,t_1)\right ),
\end{equation}
with 
$$
F(t_2,t_1)=\frac{1}{t_2-t_1}
\int^{t_2}_{t_1} dt_4 \int^{t_4}_{t_1}dt_3 (t_4-t_3)H_{int}(t_2,t_4)
H_{int}(t_3,t_1)-H_{int}(t_2,t_1). 
$$
Note that this equation determines the
operator $\tilde S_1(t_2,t_1)$ only up to some quasilocal operators $\Delta
(t_2,t_1)$, i.e., operators which do not zero only for $t_2=t$, satisfying
the condition $(t_2-t_1)\Delta(t_2,t_1)=0$. However, the operator
$\tilde S_1(t_2,t_1)$, by the definition of the operator $H_{int}(t_2,t_1)$ , 
is less singular at $t_2=t_1$ than
$H_{int}(t_2,t_1)$, and therefore cannot contain the quasilocal
operators. For Eq.(15) to have a unique solution,
the operator $H_{int}(t_2,t_1)$ must be close enough to the corresponding
solution. This means that $<\Psi_2|\int^{t}_{t_0} dt_2 \int^{t_2}_{t_0}dt_1
F(t_2,t_1)|\Psi_2>$ must rapidly enough tend to zero as $t\tend t_0$ for any
vectors $|\Psi_1>$ and $|\Psi_2>$.
In the case of Hamiltonian dynamics, when $H_{int}(t_2,t_1)$ is of the form (6),
$$
F(t_2,t_1)=(-i)^2H_I(t_2)H_I(t_1). 
$$
In this case
$$
<\Psi_2| \int^{t}_{t_0} dt_2 \int^{t_2}_{t_0}dt_1
F(t_2,t_1)|\Psi_1>=o(\tau^2) 
$$
for $\tau\tend 0$, provided $H_I(t)$ is a self
adjoint operator on the Hilbert space of physical states. It is easy to verify
that the iterative solution of Eq.(15) with $H_{int}(t_2,t_1)=-2i\delta
(t_2-t_1)H_I(t_1)$ and $F(t_2,t_1)=(-i)^2H_I(t_2)H_I(t_1)$ yields the
expression (10) for the operator $\tilde S(t_2,t_1)$.
Note that one can include the term
$(-i)^2H_I(t_2)H_I(t_1)$ into the generalized interaction operator.
The operator $H_{int}(t_2,t_1)$ constructed in this way is  closer
to the solution of Eq.(3) than $-2i\delta(t_2-t_1)H_I(t_1)$:
In this case we start with the generalized
interaction operator, being the solution of Eq.(3) in the second order
approximation. Obviously, the generalized interaction operators
$-2i\delta(t_2-t_1)H_I(t_1)$ and $-2i\delta
(t_2-t_1)H_I(t_1)+(-i)^2H_I(t_2)H_I(t_1)$ are dynamically equivalent
because they yield the same dynamics. However, the operator
$-2i\delta(t_2-t_1)H_I(t_1)$ contains the all needed dynamical information,
and it is natural to use this operator as the generalized interaction operator
generating Hamiltonian dynamics.

Let us come back to our model now. The production $(-i)^2H_I(t_2;g)
H_I(t_1;g)$ has the nonintegrable singularity at the point $t_2=t_1$, and
as a consequence is not an operator valued distribution, i.e.
$\int^{t}_{t_0} dt_2 \int^{t_2}_{t_0}dt_1(-i)^2H_I(t_2;g)
H_I(t_1;g)$ is not an operator on Fock space.
For this reason we have to introduce the momentum
cutoff that should be removed after renormalization. This means that
in Eq.(15) we have to use the generalized interaction operator
$H_{int}(t_2,t_1)=-2i\delta(t_2-t_1)H^{(r)}_I(t_1;g)$. In this case
the operator $F(t_2,t_1)$ is of the form $F(t_2,t_1)=
(-i)^2H^{(r)}_I(t_2;g)H^{(r)}_I(t_1;g)$. However we cannot let $\Lambda
\tend\infty$ in Eq.(15) with such operators $H_{int}(t_2,t_1)$ and
$F(t_2,t_1)$, since the counterterms $i\delta(t_2-t_1)\delta m^2(\Lambda)
\int d^2x g({\bf x}):\varphi^2({\bf x},t_1):$  and  $2i\delta(t_2-t_1)E(\Lambda)
\int d^2x g({\bf x})$, that have to eliminate the divergences of the operator
$\lim\limits_{\Lambda\tend\infty}(-i)^2H^{(r)}_I(t_2;g)H^{(r)}_I(t_1;g)$
in $F(t_2,t_1)$, are contained in the operator $H_{int}(t_2,t_1)$.
On the other hand, we can take the limit $\Lambda\tend\infty$ in Eq.(15),
if we start with the interaction operator 
$$
H_{int}(t_2,t_1)=-2i\delta
(t_2-t_1)H^{(r)}_I(t_1;g)+(-i)^2H^{(\Lambda)}_I(t_2;g)H^{(\Lambda)}_I(t_1;g)
$$
that is dynamically equivalent to the operator
$-2i\delta(t_2-t_1)H^{(r)}_I(t_1;g)$ for finite $\Lambda$. In fact, in
this case the generalized interaction operator $H^{(\Lambda)}_{int}(t_2,t_1)$
contains the term $(-i)^2H^{(\Lambda)}_I(t_2;g)H^{(\Lambda)}_I(t_1;g)$ together
with the corresponding counterterms $i\delta(t_2-t_1)\delta m^2(\Lambda)
\int d^2x g({\bf x}):\varphi^2({\bf x},t_1):$
and $2i\delta(t_2-t_1)E(\Lambda)
\int d^2x g({\bf x})$. However, since in the model under study $<0|H_I(t_1;g)
H_I(t_2;g)H_I(t_3;g)|0>$ has the nonintegrable singularity at the point
$t_1=t_2=t_3$, the counterterm $E(\Lambda)$ is of the form $E(\Lambda)=
E_2(\Lambda)+E_3(\Lambda)$, where $E_2(\Lambda)$ and $E_3(\Lambda)$ are
the vacuum energy counterterms in the second and third orders respectively.
For this reason we have to include the term $\int_{t_1}^{t_2}dt(-i)^3<0|
H^{(\Lambda)}_I(t_2;g)H^{(\Lambda)}_I(t;g)H^{(\Lambda)}_I(t_1;g)|0>$
into the generalized interaction operator. After this we can let
$\Lambda\tend\infty$. In this way we get the generalized interaction
operator of the theory
\begin{equation}
H_{int}(t_2,t_1)=-2i\delta(t_2-t_1)H_I(t_1;g)+H_{non}(t_2,t_1),
\end{equation}
with
$$
H_{non}(t_2,t_1)=\lim\limits_{\Lambda\tend\infty}( (-i)^2
H^{(\Lambda)}_I(t_2;g)H^{(\Lambda)}_I(t_1;g)+
$$
$$
+(-i)^3\int_{t_1}^{t_2}dt
<0|H^{(\Lambda)}_I(t_2;g)H^{(\Lambda)}_I(t;g)H^{(\Lambda)}_I(t_1;g)|0>+
i\delta(t_2-t_1)\delta m^2(\Lambda)\times
$$
$$
\times\int d^2x g({\bf x}):\varphi^2({\bf x},t_1):
+2i\delta(t_2-t_1)E(\Lambda)
\int d^2x g({\bf x})).$$
This is the generalized interaction operator of the renormalized $\varphi^4_3$
theory. This operator is an operator-valued distribution on Fock space,
i.e. there are no problems associated with the UV divergences.
Correspondingly Eq.(15) with such an operator $H_{int}(t_2,t_1)$ is
ultraviolet finite and directly leads to the renormalized  expression for
the evolution operator and S matrix. In fact, the operator (16) is defined as
a limit of the consequence of the operators $H^{(\Lambda)}_{int}(t_2,t_1)$
which are dynamically equivalent to the operators
$-2i\delta(t_2-t_1)H^{(r)}_I(t_1;g)$, and generates Hamiltonian dynamics.
The perturbative solution of Eq.(15) for these interaction operators are
known, and are given by (11). Thus the perturbative solution of Eq.(15) with
the generalized interaction operator (16) is a limit of the consequence
of the operators $\tilde S_{r,\Lambda}(t_2,t_1;g)$ for $\Lambda\tend\infty$.
From this and (14) it follows that the solution of Eq.(15) with the generalized
interaction operator (16) yields the same results as the renormalized
theory $\varphi_3^4$. Note in this connection that the renormalization
program implies that all observable quantities, such as S matrix, can
be represented as a limit of these quantities calculated by using
Hamiltonians with an approximate, momentum cutoff. The UV
divergences manifest themselves in the fact that limiting interaction
Hamiltonians are infinite and therefore physically meaningless. In fact,
the consequence of the operators $H^{(r)}_I(t_1;g)$
does not converge to some finite operator. At the same time the consequence
of the operators $H^{(\Lambda)}_{int}(t_2,t_1)$, that are dynamically
equivalent to the operators $-2i\delta(t_2-t_1) H^{(r)}_I(t_1;g)$ for
finite $\Lambda$, converges to some operator valued distribution
on Fock space, that is not dynamically equivalent
to any generalized interaction operator of the form (6), i.e. is
nonlocal in time. Thus the dynamics of the renormalized theory is
generated by the nonlocal interaction being described by the
generalized interaction operator (16).

In summary, we have shown that after renormalization the dynamics
of the theory $\varphi^4_3$ is governed by the generalized
dynamical equation (3) with the nonlocal interaction operator (16).
By solving this equation with the interaction operator (16)
and removing the spartial cutoff, one can construct the finite,
and Lorents invariant S matrix. Thus within the GQD this theory
manifests itself as a finite theory free from UV divergences.
This gives reason to suppose that such a dynamical situation
takes place in any renormalizable theory, and 
such theories as QED and QCD may be formulated in an ultraviolet-finite
way. It is hoped that the above ideas may be also applied to
nonrenormalizable theories. The results of our paper also
show that the GQD may open new possibilities for applying the
EFT approach to nuclear physics. In fact, the essential lesson
we have learned from our analysis is that the low-energy
nucleon dynamics to which a EFT leads after renormalization 
should be governed by Eq.(3) with nonlocal generalized
interaction operator that provides a natural parameterization
of the nuclear forces. By using a EFT, one can construct a
generalized operator of the NN interaction consistent with symmetries of QCD.
This operator can then be used in Eq.(3) for describing nucleon dynamics.

The work was supported by Fund NIOKR of RT and Academy of Sciences of
Republic of Tatarstan N 14-98.

\newpage
\section*{References}

\begin{enumerate}

\item[{[1]}]
D. Evens, J.W. Moffat, G. Kleppe, and B.P. Woodard, Phys. Rev. D {\bf 43},
499 (1991); N.J. Cornish, Int. J. Mod. Phys. A {\bf 7}, 6121 (1992);
G.V. Efimov, Nonlocal Interactions of Quantized Fields (Nauka, Moscow, 1977).
\item[{[2]}]
R.Kh. Gainutdinov, J. Phys. A: Math. Gen. {\bf 32}, 5657 (1999).
\item[{[3]}]
S.Weinberg, Phys. Lett. B {\bf 251}, 288 (1990); Nucl. Phys. B {\bf 363},
3 (1991);C. Ordonez, and U. van Kolck, Phys. Lett. B {\bf 291}, 459 (1992);
U. van Kolck, Phys. Rev. C {\bf 49}, 2932 (1994).
\item[{[4]}]
U. van Kolck, Prog. Part. Nucl. Phys. {\bf43} 409 (1999).  
\item[{[5]}]
R. Kh. Gainutdinov and A.A. Mutygullina, nucl-th/0107003.
\item[{[6]}]
D.R. Phillips, I.R. Afnan, and Henry-Edvards, Phys. Rev. C {\bf 61} 044002 
(2000).
\item[{7]}]
R.P. Feynman, Rev. Mod. Phys. {\bf 20}, 367 (1948);
 R.P. Feynman  and A.R. Hibbs, Quantum Mechanics and Path
  Integrals, (McGraw-Hill, New York, 1965).
\item[{[8]}]
R. Kh. Gainutdinov and  A.A.Mutygullina,
 Yad. Fiz. {\bf 60}, 938 (1997) [Physics of Atomic Nuclei,
{\bf 60}, 841 (1997)];
 Yad. Fiz. {\bf 62}, 2061 (1999) [Physics of Atomic Nuclei,
{\bf 62}, 1905 (1999)].
\item[{[9]}]
J. Glimm and A. Jaffe, Fortshritte der Physik {\bf 21}, 327 (1973).
\end{enumerate}

\end{document}